\documentclass{aa}
\usepackage{graphicx}
\usepackage{subfig}
\usepackage{natbib}
\usepackage{rotating}
\usepackage{multirow}
\usepackage[varg]{txfonts}

\newcommand{\xmm}{{\small \it XMM-Newton}}

\newcommand{\rgs}{{\small RGS}}
\newcommand{\chandra}{{\small \it Chandra}}
\newcommand{\swift}{{\small {\it Swift}}}
\newcommand{\nustar}{{\small {\it NuSTAR}}}

\newcommand{\hst}{{\small {\it HST}}}

\newcommand{\ngc}{{\small NGC\,7469}}

\newcommand{\kms}{km\,s$^{-1}$}
\newcommand{\ergs}{erg\,s$^{-1}$}
\newcommand{\cmsq}{cm$^{-2}$}
\newcommand{\flux}{erg\,s$^{-1}$cm$^{-2}$}
\newcommand{\ionpar}{erg\,s$^{-1}$cm}

\newcommand\nodata{ ~$\cdots$~ }

\begin{document}

\title{Multi-wavelength campaign on NGC\,7469 I. \\ The rich 640 ks RGS spectrum}
\titlerunning{Multi-wavelength campaign on NGC\,7469}
\authorrunning{Behar et~al.}


\author{Ehud Behar\inst{1,2}, Uria Peretz\inst{1}, Gerard A. Kriss\inst{3}, Jelle Kaastra\inst{4,5}, Nahum Arav\inst{6,1}, Stefano Bianchi\inst{7}, Graziella Branduardi-Raymont\inst{8}, Massimo Cappi\inst{9},  Elisa Costantini\inst{4}, Barbara De Marco\inst{10}, Laura Di Gesu\inst{11}, Jacobo Ebrero\inst{12}, Shai Kaspi\inst{13}, Missagh Mehdipour\inst{4},  St\'{e}phane Paltani\inst{11}, Pierre-Olivier Petrucci\inst{14}, Gabriele Ponti\inst{10}, Francesco Ursini\inst{14}
}
\institute{Department of Physics, Technion, Haifa 32000, Israel \and 
Department of Astronomy, University of Maryland, College Park, USA \and 
Space Telescope Science Institute, 3700 San Martin Drive, Baltimore, MD 21218, USA \and 
SRON Netherlands Institute for Space Research, Sorbonnelaan 2, 3584 CA Utrecht, The Netherlands \and 
Leiden Observatory, Leiden University, PO Box 9513, 2300 RA Leiden, The Netherlands \and
Department of Physics, Virginia Tech, Blacksburg, VA 24061, USA \and 
Dipartimento di Matematica e Fisica, Universit\`{a} degli Studi Roma Tre, via della Vasca Navale 84, 00146 Roma, Italy \and
Mullard Space Science Laboratory, University College London, Holmbury St. Mary, Dorking, Surrey, RH5 6NT, UK \and
INAF-IASF Bologna, via Gobetti 101, 40129 Bologna, Italy \and 
Max-Planck-Institut f\"{u}r extraterrestrische Physik, Giessenbachstrasse, 85748 Garching, Germany \and
Department of Astronomy, University of Geneva, 16 Chemin d'Ecogia, 1290 Versoix, Switzerland \and
European Space Astronomy Centre, PO Box 78, 28691 Villanueva de la Ca\~{n}ada, Madrid, Spain \and
School of Physics and Astronomy, Tel Aviv University, Israel \and
Univ. Grenoble Alpes, CNRS, IPAG, F-38000 Grenoble, France 
}

\abstract{}
{Outflows in active galaxies (AGNs) are common, although their launching mechanism, location, and physical impact on the host galaxy remain controversial. We conducted a multiwavelength six-month campaign to observe the nearby Seyfert galaxy \ngc\ with several observatories in order to better understand and quantify the outflow in this AGN.}
{We report on the time-integrated line-resolved X-ray spectrum of \ngc\ obtained with the Reflection Grating Spectrometer (\rgs ) on board \xmm . 
We use the \rgs\ spectrum to discern the many AGN outflow components. 
A global fit is applied to obtain their physical parameters. }
{We find that the AGN wind can be well described by three narrow velocity components at $\sim$ --650, --950, and --2050 \kms. 
The \rgs\ clearly resolves the --2050\,\kms\ component in C$^{5+}$ Ly\,$\alpha$, while the --650\,\kms\  and --950\,\kms\  velocities are blended.
Similar velocities ($\pm\ 200$\,\kms ) are resolved in the UV.
The H-equivalent column densities of these components are, respectively, $N_{\rm H} \sim 7\times 10^{20}$, $2.2\times 10^{21}$, and $10^{20}$ \cmsq , for a total of $\sim 3\times 10^{21}$ \cmsq , which was also measured in 2004, indicating the absorber did not significantly change.
The --650\,\kms\ component shows a broad ionisation distribution ($-1 \lesssim \log \xi \lesssim 2,~\xi $ being the ionisation parameter in \ionpar).
We identify a photo-ionised emission component blue-shifted by $\sim$ --450\,\kms , somewhat broad (FWHM = 1400\,\kms ), and with $-1 \lesssim \log \xi \lesssim 1$ \ionpar , which we ascribe to the same outflow that produces the absorption lines.
We also find a collisionally-ionised component at $kT = 0.35$\,keV that we associate with the  circum-nuclear star formation activity of \ngc , as it follows the $L_{\rm FIR} / L_{\rm X} \approx 10^4$ relation found in star forming galaxies.
The elemental abundance ratios  of C, N, Ne, S, and Fe to O in the outflow tend to be between 1 - 2 times solar.
Preliminary estimates of the absorber distance from the AGN center suggest it is at least a few pc away from the center, but more advanced methods need to be applied in order to obtain better constraints.}
{The complex X-ray spectrum of \ngc\ demonstrates the richness of high energy phenomena taking place in AGN cores. 
The subtle spectroscopic differences between the various components require deep, high-resolution observations, such as the present \rgs\ spectrum, if one is to resolve them and perform quantitative plasma diagnostics.}


\maketitle

\section{Introduction}
\label{intro}

AGN outflows couple supermassive black holes to their environments \citep[e.g.,][]{DiMatteo05}. 
Despite many years of spectroscopic observations in the X-rays and in the UV, the emerging picture of these outflows is still ambiguous.
Spectra are typically characterized by several kinematic and ionisation components that are readily measured with contemporary instruments \citep[see][for a review]{Crenshaw03}. 
However, the relation between the various components remains debatable.
Reaching a more complete physical picture of AGN outflows has proven to be challenging \citep[e.g.,][]{Blustin05}.
Thus, their impact, perhaps most interestingly their mass outflow rate that affects their momentum and energy feedback potential to the host galaxy remains largely unknown.

Seyfert outflows for the most part eject winds of a few 100\,\kms , and with $v/c < 0.01$ where $c$ is the speed of light \citep{Kaastra00, Kaspi01}. 
Their absorption spectra are extremely rich and can feature essentially all charge states from neutral to H-like ions. These winds thus appear to be different from the ultra high velocity outflows ($v/c \sim 0.1$) observed mostly through their  Fe$^{25+}$ absorption \citep{Reeves03, Tombesi10}. Since the mechanical power in AGN outflows scales with the mass outflow rate and with $v^2$, and since the luminosity $L$ of an AGN scales with the mass accretion rate and $c^2$, mass outflow rates in Seyferts need to be 10,000 times higher than the accretion rate for their mechanical power to reach the AGN luminosity.
On the other hand, the high-resolution high-S/N spectra observed in Seyferts serve as fertile grounds for studying the physics of outflows from black hole accretion sources in the greatest detail possible.   

The theoretical models of AGN outflows vary greatly.
The launching mechanism remains controversial, as thermal evaporation of the torus \citep[e.g.,][]{Krolik01}, radiation pressure  
\citep[e.g.,][]{Proga00}, and magnetohydrodynamics driving \citep[e.g.,][]{Contopoulos94} have all been proposed.
Nevertheless, matching observed spectra with a real physical model has proven to be difficult \citep[e.g.,][]{Sim08}.
Very different physical models have been successful in producing the broad ionisation distribution, both with MHD winds \citep{Fukumura10}, and with radiation compression \citep{Rozanska06, Stern14}. 
The kinematic structure obtained when fitting for discrete velocity and ionisation components in each AGN remains harder to explain from first principles, as those do not seem to follow any obvious trend \citep{Blustin05, McKernan07}.
\citet{Laha14} find interesting trends between targets of outflow velocity and column density that increase with ionisation parameter, which could represent an important clue to the physical mechanism of these winds.

The mass outflow rate depends on the location (and density) of the ionised absorber.
Since X-ray imaging does not resolve line-of-sight absorbers, several methods have been proposed and used to estimate the distance from the center ($r$) or the H density ($n_{\rm H}$) and thus to lift the hampering degeneracy between the two, which is imposed by the measured ionisation parameter $\xi = L / (n_{\rm H}r^2)$. 
The most accurate of these methods is identification of absorption from density-sensitive metastable levels \citep{Arav01, Arav15}.
However, in the X-rays, most such levels require high densities that are not observed, and low-density diagnostics have proven to be extremely complicated to implement \citep{Kaastra04}.
Associating the absorber with the narrow emission lines is yet another method. 
Fluxes of lines driven by recombination scale with $n_{\rm H}^2$, while the column density scales with $n_{\rm H}$, which allows to obtain the density \citep[e.g.,][]{Behar03}. However,  this method works only when there are bright emission lines.
Another common approach is to estimate recombination or ionisation times as the absorber responds to changes in the ionising flux.
In many cases, the ionisation of the absorber does not change on short time scales, despite variability of the continuum \citep{Behar03, Netzer03, Ebrero10, Kaastra12}, which constrains it to be many pc away from the nucleus, and the density to be low.
Other works do suggest a response of the absorber ionisation to the continuum level that places the absorber closer than 6\,pc in {\small NGC\,3783} \citep{Krongold05}, and as close as a few light days in {\small NGC\,4051} \citep{Krongold07}.
For the most part, non-continuous monitoring impedes transient ionisation/recombination modeling, and one needs to revert to statistical correlations between flux level and ionization state \citep{Krongold07}, or column density of certain ions \citep{Kaastra12}.

The present campaign aims to harness \xmm , \hst , \chandra , \nustar , and \swift , as well as ground based telescopes, to monitor the changes of the continuum and the response of the ionised absorber of \ngc .
Similar successful campaigns in the past include other Seyfert galaxies, such as {\small NGC\,3783} \citep{Netzer03},  {\small Mrk\,279} \citep{Gabel05, Ebrero10}, {\small Mrk\,509} \citep{Detmers11,Kaastra12}, and {\small NGC\,5548} \citep{Kaastra14}.

The unique properties of \ngc\ make it the ideal candidate to respond on shorter, better-monitored time scales.
\ngc\ is as luminous as {\small Mrk\,509} ($L_{\rm bol} \approx 10^{45}$erg\,s$^{-1}$),
but hosts a black hole that is 10 times smaller ($M_{\rm BH}$ = $10^7 M_\odot$),
resulting in a high Eddington ratio of $L/L_{\rm Edd} \approx 0.3$.
Given this compactness, one expects 10 times faster variability.
Indeed, the normalised excess variance of \ngc\ is ten times higher on short time scales of $< 100$\,ks than that of {\small Mrk\,509} \citep{Ponti12}.
In the 0.3 - 10~keV band, \ngc\ is one of the ten brightest Seyfert~1 galaxies with a flux of $4-5\times 10^{-11}$erg\,s$^{-1}$cm$^{-2}$.

Previous analyses of X-ray \chandra\ and \xmm\ grating spectra of  \ngc\ were published by \citet{Scott05} and by \citet{Blustin07}, respectively. \citet{Blustin07} found a wide range of ionisation, with $-0.5 < \log \xi < 3.5$ (\ionpar ), and two main velocity regimes, at $\sim -650$ and --2300 \kms , the former of which was also identified by \citet{Scott05}.
The total absorbing column density in the X-rays was of order $3\times 10^{21}$ \cmsq .
The previous works also identified the lowest ionisation phase of the X-ray absorber with one of the phases of the UV absorber. 
They estimated the outflow to be at the base of the torus.
In this paper, we revisit the ionised absorber of \ngc\ and focus on the co-added 640~ks spectrum, acquired with seven separate observations of the Reflection Grating Spectrometer (\rgs ) on board \xmm . 
The time dependence as well as the simultaneous UV and hard X-ray spectra will be discussed in future companion papers.

We take the rest frame of \ngc\ to be at a redshift of z=0.016268, based on the 21cm line \citep{Springob05}, and the luminosity distance to be 65.1 Mpc.
For the total Galactic column density towards \ngc , we add the H\,I column of $4.34\times 10^{20}$\cmsq\ \citep{Wakker11} and  twice the measured H$_2$ column density of $5.75\times 10^{19}$\cmsq\ \citep{Wakker06}, for a total H column of $5.5\times 10^{20}$\cmsq . 
For the solar abundances we refer to \citet{Asplund09}. 
The remainder of the paper includes the data reduction (Sec.~\ref{data}), the fitting method (Sec.~\ref{method}), the results  (Sec.~\ref{results}), and discussion (Sec.~\ref{discussion}).

\section{Data and Reduction}
\label{data}

\xmm\ observed \ngc\ seven times for approximately 90\,ks on average each time, and with varying spacings between observations,
in order to probe variability on different time scales.
Luckily, no major background flares occurred during these observations, which allowed us to essentially make use of the entire exposure times.
The observation log is given in Table~\ref{log}, including the mean \rgs 1 source count rates of each observation.
It can be seen that in terms of the \rgs\ count rate, the source varied by up to 30\% between observations.
No obvious change in the absorption features was identified. 
If they did vary, these were subtle changes that will be investigated in a separate paper.
This allowed us to co-add all \rgs\ (both 1 and 2) spectra and to study the mean \rgs\ features at a high S/N ratio.
We did verify that the global model fitted to the mean \rgs\ spectrum, as discussed below, fits each individual spectrum well.
In statistical terms, after merely normalizing the model continuum, the $\chi ^2$/d.o.f. remains   
between 1.30 to 1.57 for all individual spectra, while the best fit model to the combined spectrum yields $\chi ^2$/d.o.f. = 1.40.

\begin{table}
\caption{Observation Log }            
\label{log}      
\centering                                      
\begin{tabular}{c c c c}          
\hline\hline                        
Observation & Start date & Duration & \rgs 1 count rate \\    
id &  & (ks) & (cts/s) \\
\hline                                   
0760350201 & 2015-Jun-12 & 89.5 & 0.685  \\ 
0760350301 & 2015-Nov-24 & 85.6  &  0.600 \\ 
0760350401 & 2015-Dec-15 & 84.6  & 0.521  \\ 
0760350501 & 2015-Dec-23 & 89.5  & 0.536  \\ 
0760350601 & 2015-Dec-24 & 91.5 & 0.593  \\ 
0760350701 & 2015-Dec-26 & 96.7 & 0.580  \\ 
0760350801 & 2015-Dec-28 & 100.2 & 0.620  \\
\hline                                             
\end{tabular}
\end{table}

We reduced each \rgs\ (1 and 2) spectrum with the 'rgsproc' task of the standard XMM/SAS pipeline tool version 14 (xmmsas\_20141104\_1833).
The default spectral binning presented in this work is of 20\,m\AA , which over-samples the instrumental line spread function by a factor of a few.
All seven \rgs\ spectra were subsequently combined using the 'rgscombine' task.
The present paper focuses on this combined, mean spectrum.
Each \rgs\ observation had a slightly different pointing. 
This was done on purpose to dither the chip gaps on the \rgs\ detectors.
Indeed, the combined fluxed spectrum has no striking gaps, with some residuals just barely noticed as narrow jumps in the spectrum at 9.9, 13.1, 20.75, and 22.7~\AA .
The mean energy and photon flux measured with the combined (mean) \rgs\ spectrum between 0.3 - 2.5~keV is 3.15$\times 10^{-11}$\,\flux , and 0.027 ph\,s$^{-1}$cm$^{-2}$, respectively.

\section{Fitting Approach}
\label{method}

We conducted a global fit to the total \rgs\ spectrum between $7 - 38$ \AA\ using the {\small Xspec11}\footnote{http://heasarc.nasa.gov/xanadu/xspec/} software package.
Both the C and $\chi ^2$ statistic for fitting were tested.
Given the high S/N ratio, they yield essentially the same results.
From here on, we refer only to the C statistic.
We started with a Galactically absorbed 
power-law continuum.
We also tested whether a soft-excess (e.g., black body) component improves the fit.
Within the limited \rgs\ band it does not, but a broader perspective that includes hard X-rays and UV does require such a component (Mehdipour et~al., in preparation) and will be modeled properly in a subsequent paper (Petrucci et~al., in preparation). 

We subsequently use the analytical version (2.1ln8) of the {\small XSTAR} photo-ionised plasma models \citep{Kallman01}.
These models employ pre-calculated photo-ionisation balance results, using a template power-law spectrum with a photon index of $\Gamma = 2$, so they do not fully reflect the actual broad-band spectrum of each source.
Nonetheless, the present best-fit model, limited to the \rgs\ band, has $\Gamma = 2.17$ and no soft excess, so we expect this to be a good approximation.
We find that six ({\it warmabs}) photo-ionised absorption components, and two ({\it photemis}) emission components provide a satisfactory description of the spectrum. 
Since we identify ionised absorption at the observed frame, likely due to local hot gas, we also add a {\it hotabs} component at $z = 0$. 
For more information about the models, we refer the reader to the {\small XSTAR} website\footnote{https://heasarc.gsfc.nasa.gov/xstar/xstar.html}.
Finally, residuals of Fe L-shell emission lines motivated adding a collisionally-ionised component.

Simultaneous UV spectra of \ngc\ reveal three main outflow velocity components at --540, --857, and --1865~\kms\ that are roughly 60, 35, and 55 \kms\ ($\sigma$) wide, and a fourth weak one at --1400\,\kms .
We used the three main UV velocity shifts and their broadening as a starting point for  the corresponding parameters of the X-ray absorber model fit, and added ionisation components at each velocity as required by the data.
However, it is important to keep in mind that the spectral resolution of the \rgs\ is roughly $\Delta \lambda = 70$\,m\AA\ across the band. 
Consequently, its kinematic resolving power around $\lambda = 20$\AA\ is approximately $v = (\Delta \lambda / \lambda) c \approx 1000$ \kms .

We aim to measure both the column density $N_{\rm H}$ of each absorption component and the abundances of those elements that feature strong absorption and emission lines. 
Since no discrete H features appear in the X-ray spectrum, our result for $N_{\rm H}$ scales with the overall metallicity.
Thus, in the model $N_{\rm H}$ can be degenerate with the metal abundances. 
We therefore froze the abundance of O that has many features in the spectrum to its solar value, and fitted for $N_{\rm H}$ and for the other abundances.
The abundances in all absorption and photo-ionised emission components were tied together, assuming they are all the same.
The abundances of the local absorption components (neutral and ionised) were kept at their solar values.

\section{Results}
\label{results}

\begin{figure*}
\begin{center}
\includegraphics[scale=0.55,angle=270]{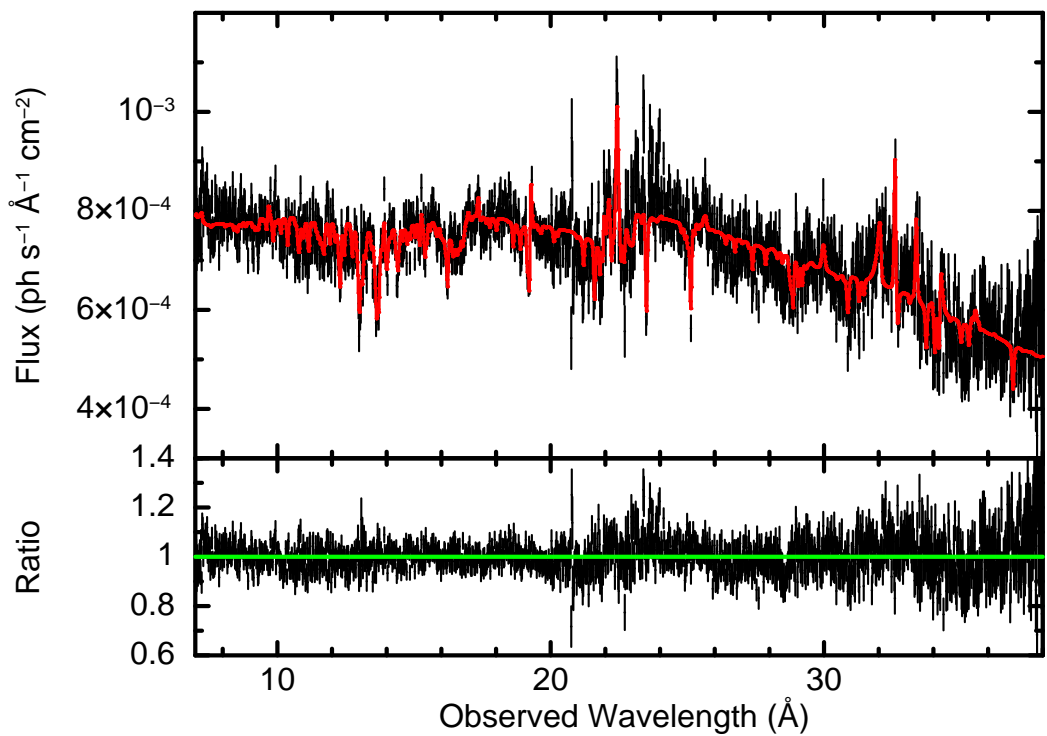} 
\end{center}
\caption{
Total \rgs\ spectrum plotted with 20\,m\AA\ bins, and the best-fit model overlaid, with the data/model ratio in the lower panel. }
\label{total}
\end{figure*}

\begin{figure*}
\begin{center}
\includegraphics[scale=0.55,angle=270]{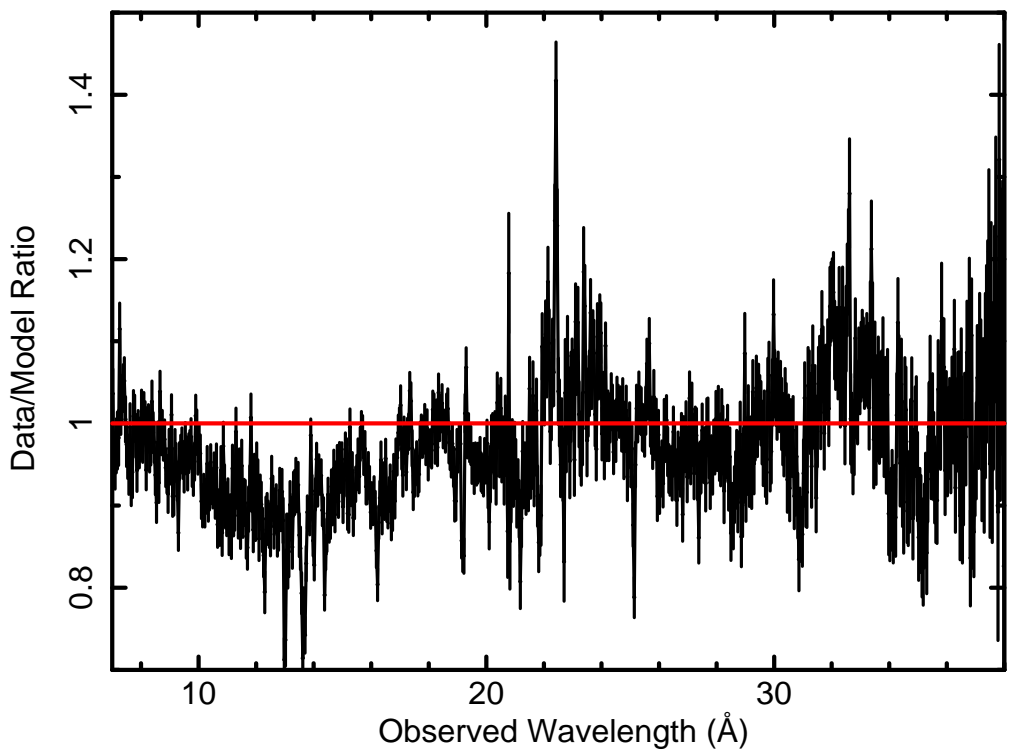} 
\end{center}
\caption{
Ratio of \rgs\ spectrum to Galactically absorbed continuum model plotted with 40\,m\AA\ bins, demonstrating the effect of ionised absorption and emission. }
\label{cont}
\end{figure*}

The data and best-fit model are plotted in Fig.~\ref{total}.
It can be seen that the overall fit is good, albeit some remaining residuals around $\sim$23 \AA , and long-ward of 30~\AA .
The best-fit model yields a C statistic of 2151 (or $\chi ^2 =$ 2179) with 1550 spectral bins and 1528 degrees of freedom.
The best-fit parameters are presented and discussed in the following sections. 
In Fig.~\ref{cont} we plot the ratio of the data to the Galactically absorbed continuum, in order to demonstrate the ionised absorption and emission. 
The relatively shallow absorption ($\sim 20\%$) and bright emission lines can be easily discerned.

\begin{figure*}
\begin{center}
\includegraphics[scale=0.37,angle=270]{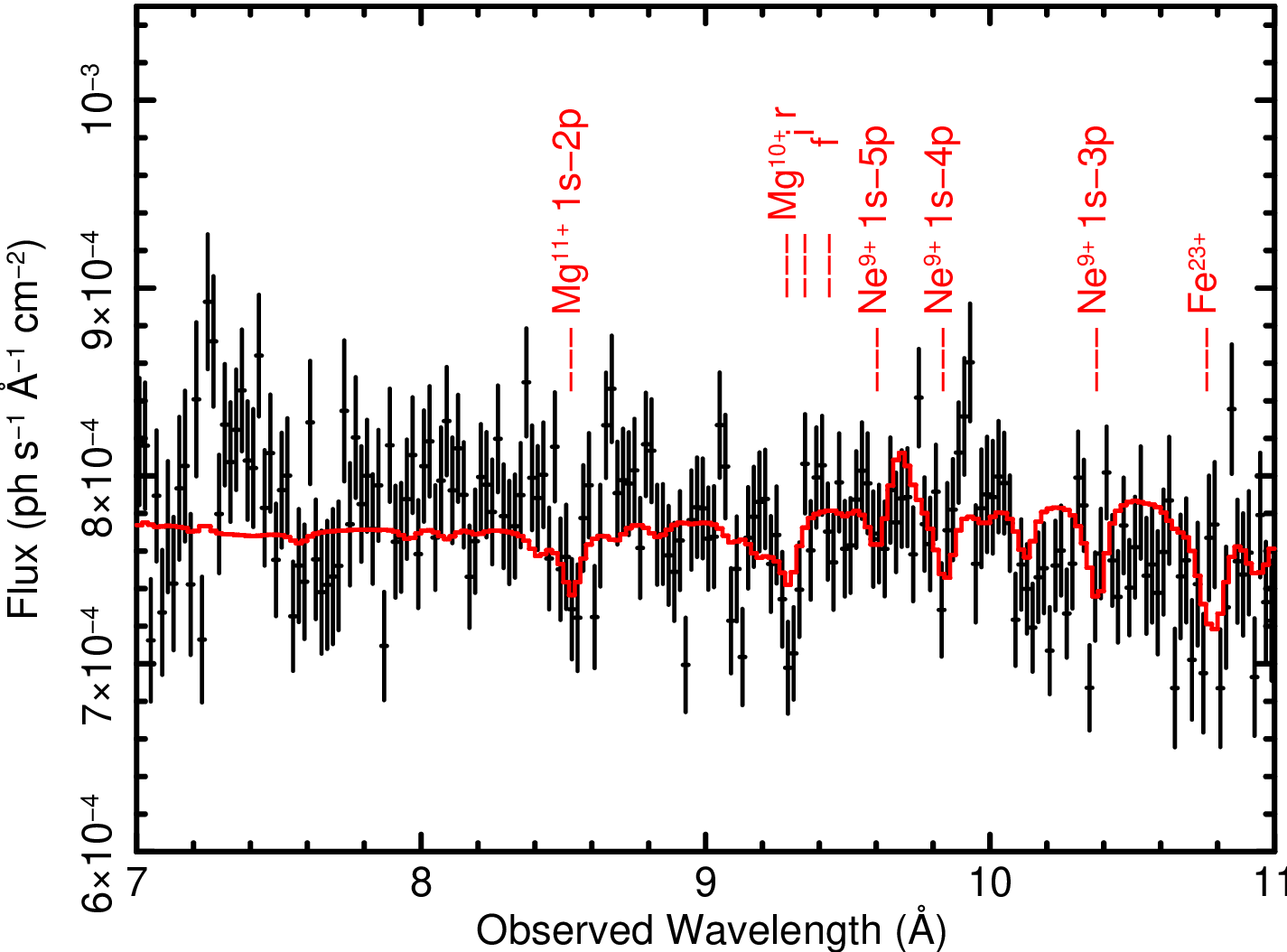} 
\includegraphics[scale=0.37,angle=270]{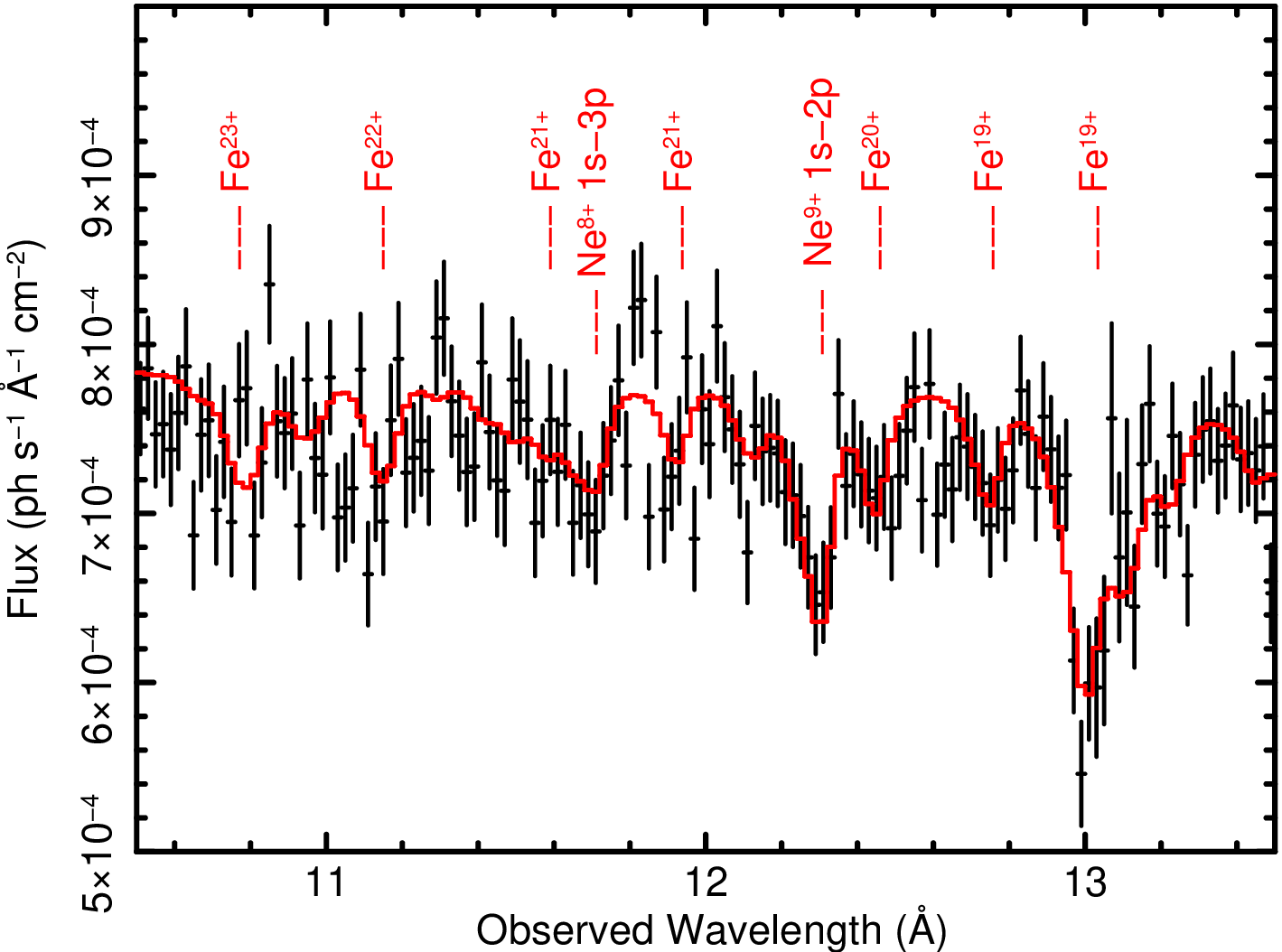} 
\includegraphics[scale=0.37,angle=270]{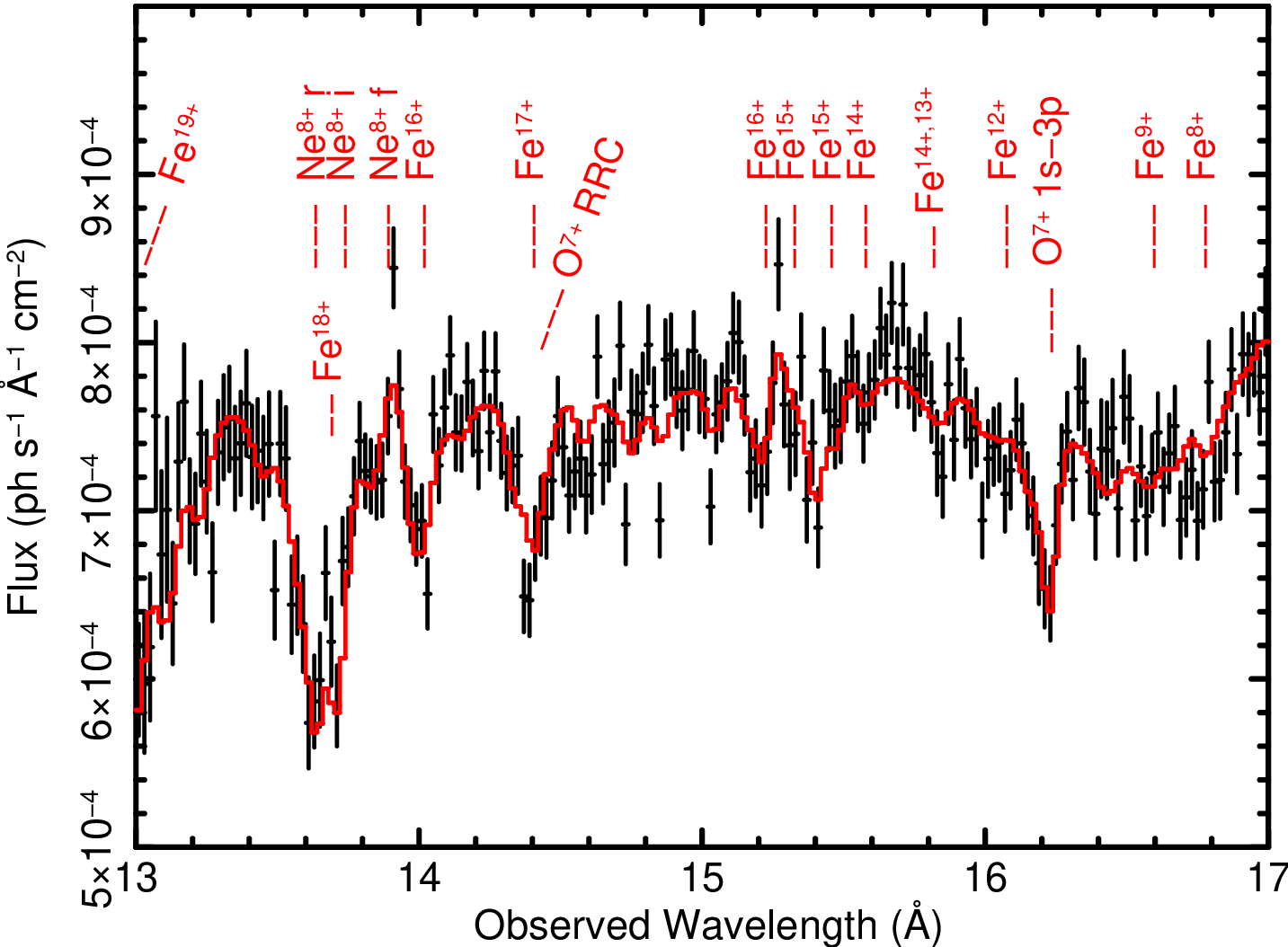} 
\includegraphics[scale=0.37,angle=270]{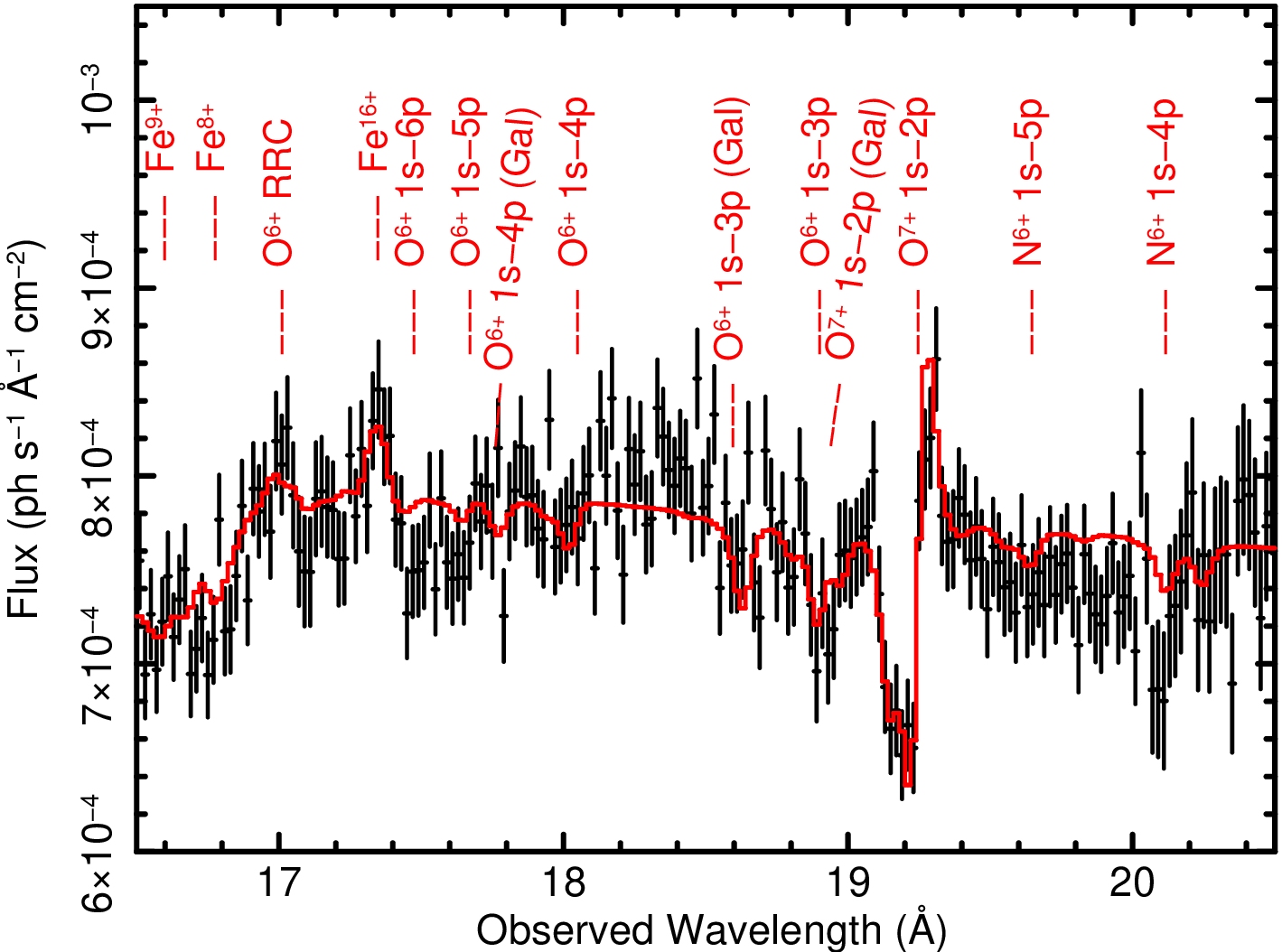} 
\end{center}
\caption{
Segments of the \rgs\ spectrum of \ngc\ with the best-fit folded model overlaid.
Spectra are presented in the observed (redshifted) frame.
Prominent features are marked on the spectrum at their position in the rest frame of \ngc .
Note the varying vertical scale from one panel to the other, none of which reaches zero.
Longer wavelengths presented in Fig.~\ref{zoom2} below.
}
\label{zoom}
\end{figure*}

\begin{figure*}
\begin{center}
\includegraphics[scale=0.37,angle=270]{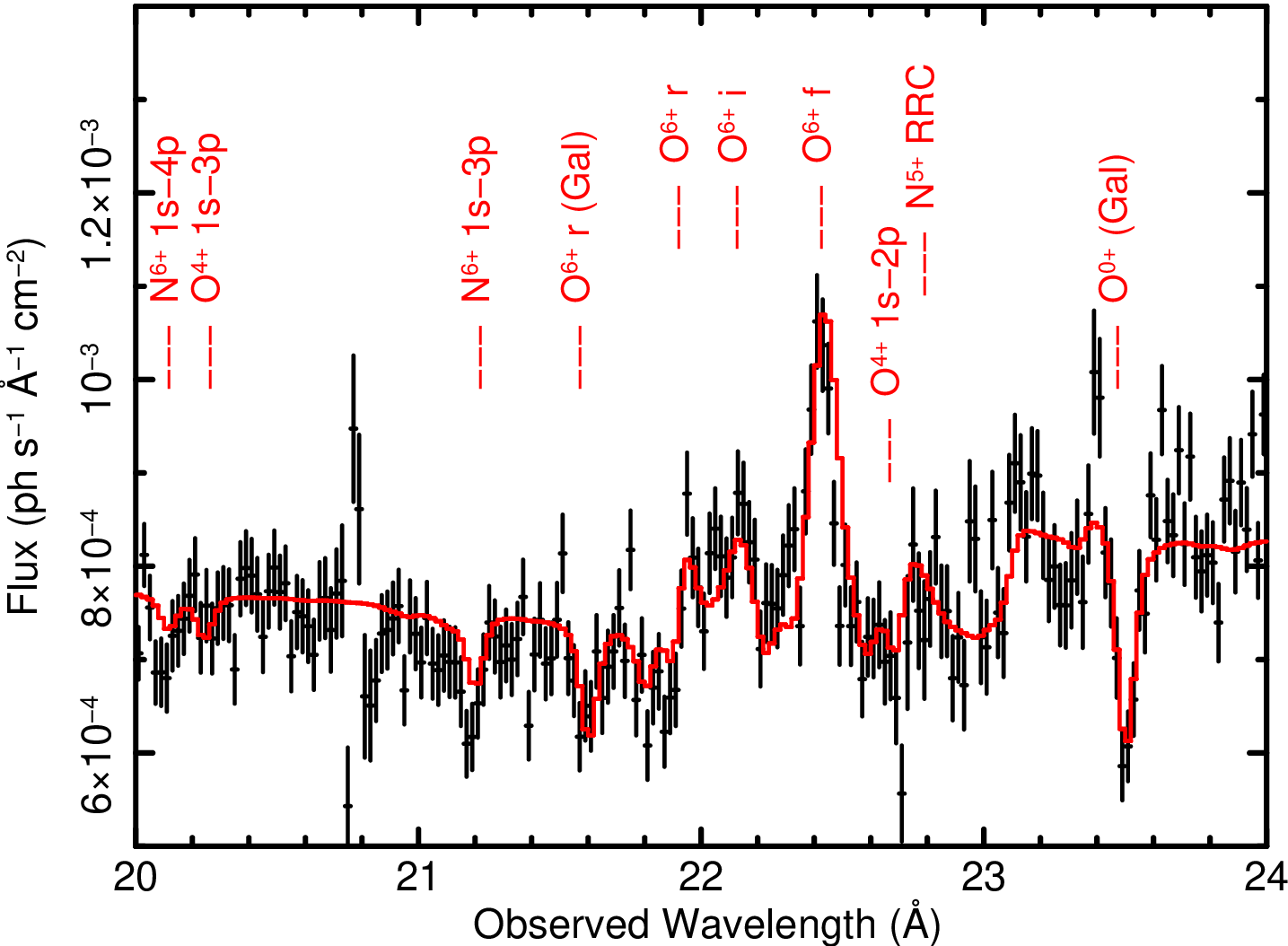} 
\includegraphics[scale=0.37,angle=270]{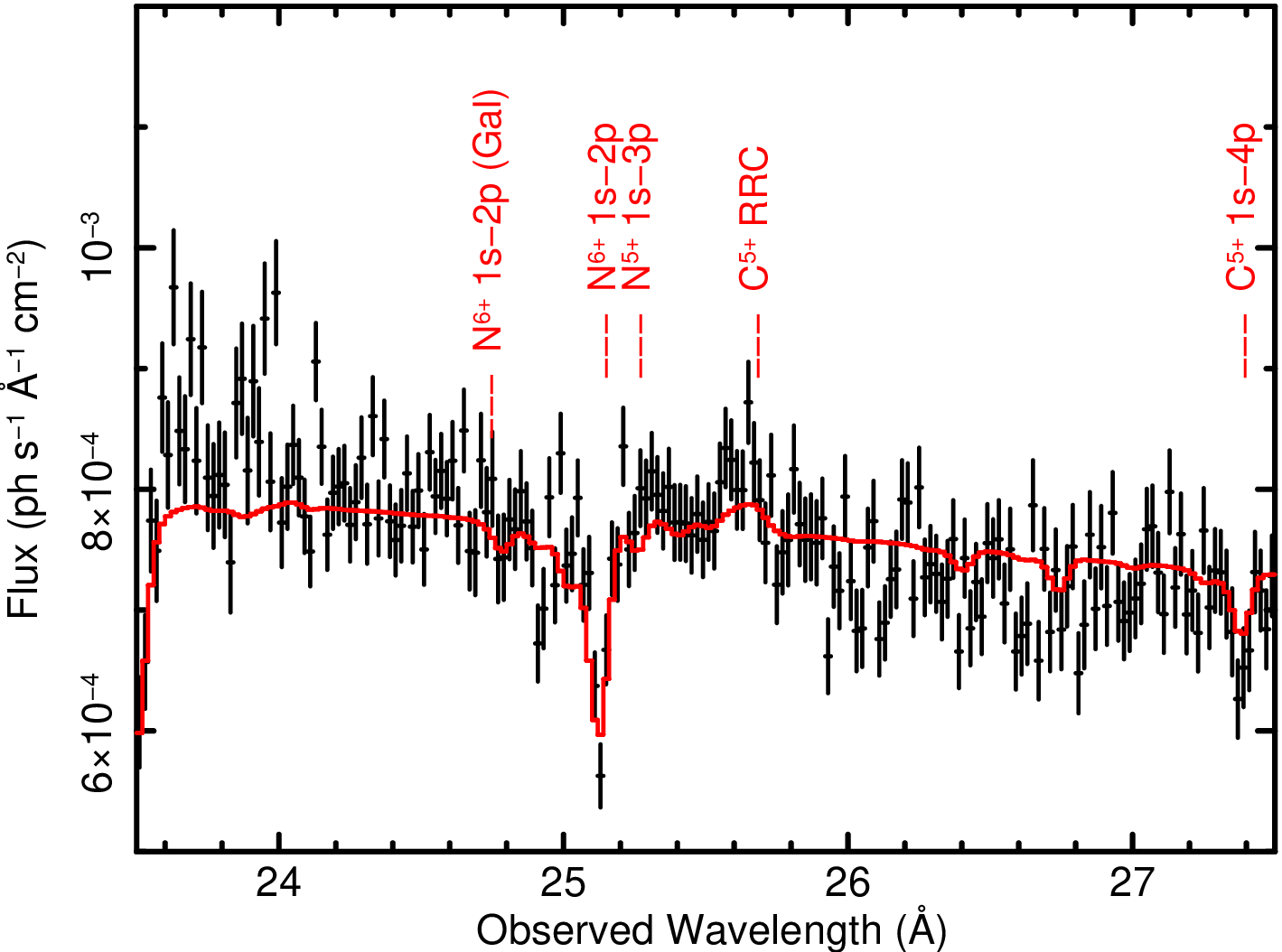} 
\includegraphics[scale=0.37,angle=270]{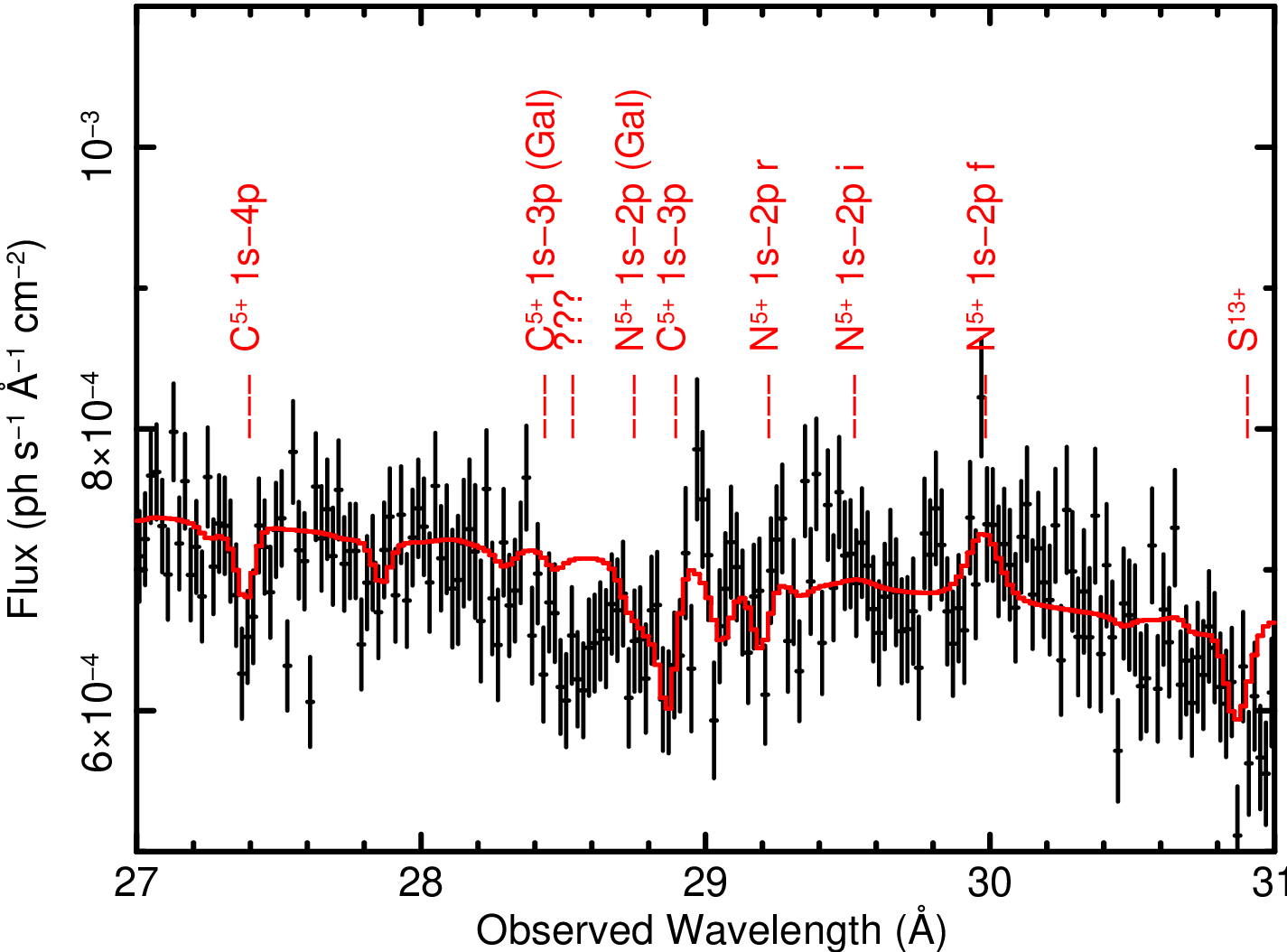} 
\includegraphics[scale=0.37,angle=270]{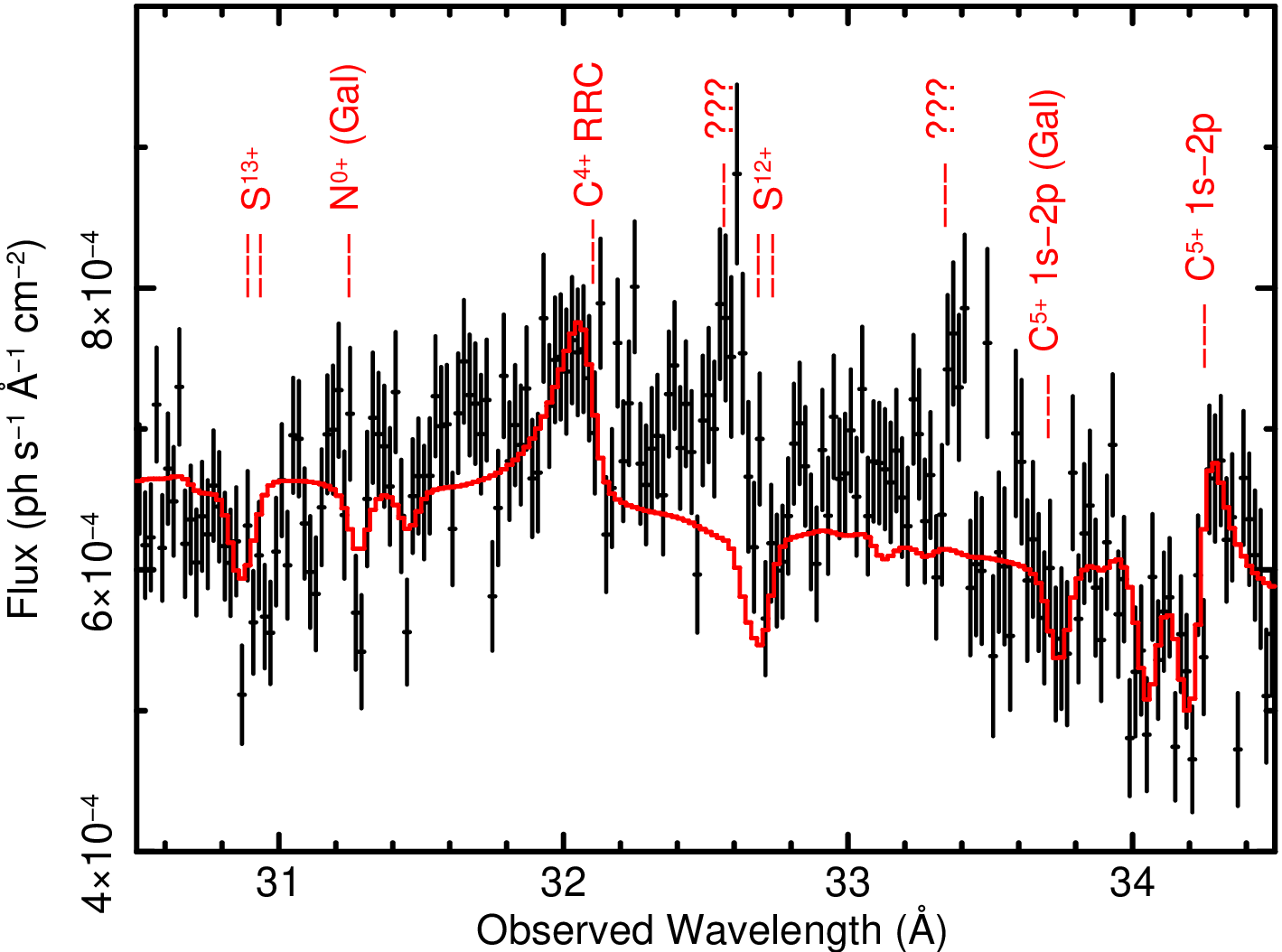} 
\includegraphics[scale=0.37,angle=270]{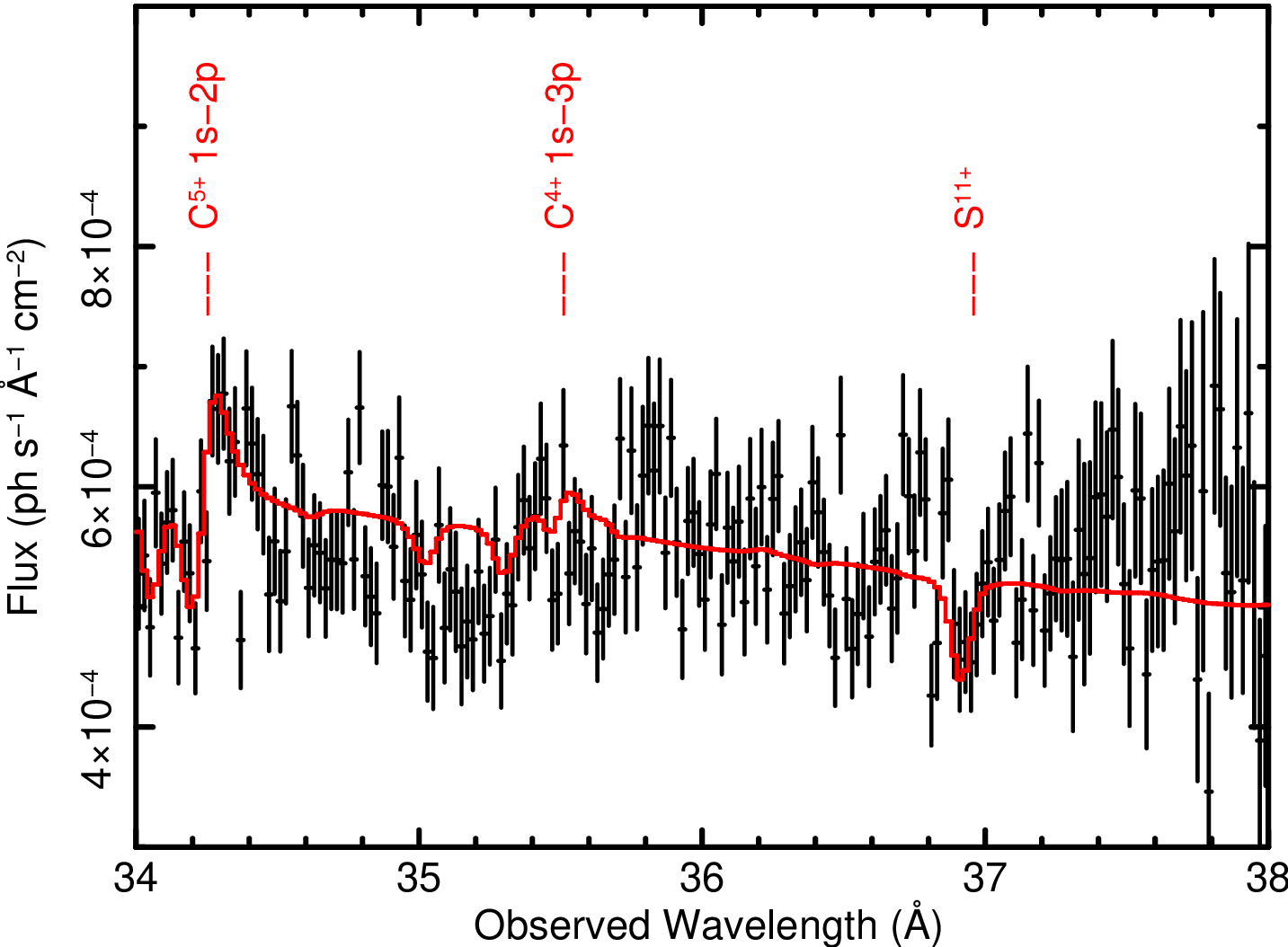} 
\end{center}
\caption{
Same as Fig.~\ref{zoom}, but for wavelengths longer than 20\,\AA .
}
\label{zoom2}
\end{figure*}

In Figs.~\ref{zoom} and \ref{zoom2} we zoom into the different regions of the spectrum to highlight the identified features. 
Here too, the fit can be seen to capture most features accurately. 
A notable exception is around the neutral O edge at $\sim$23~\AA , where either the continuum seems to be too low. 
The residuals above 30~\AA\ can be attributed predominantly to two unidentified emission lines at 32.6~\AA\ and at 33.4~\AA , which we discuss below in Sec. \ref{unidentified} in more detail.
These shortcomings of the model should have a negligible effect, if any, on the properties of the absorber (and narrow line emitter) that are the main focus of the paper.


\subsection{Kinematics}

Using three components (as in the UV), we obtain best-fit velocities of $-650 \pm 50$~\kms , $-950^{+50}_{-100}$~\kms , and $-2050 ^{+50}_{-160}$~\kms .
These velocities are slightly (100 - 200\,\kms ) higher than the UV values, but consistent within the measurement and wavelength calibration ($< 10$~m\AA ) uncertainties.
Forcing the velocities to the UV values worsens the fit by $\Delta C \approx 20$ for more than 1500 degrees of freedom.
They are also in line with previous X-ray measurements \citep{Scott05, Blustin07}, hinting that they did not vary over more than a decade.
Perhaps the best absorption (and emission) lines in the present spectrum for kinematic studies are the Ly\,$\alpha$ lines (close doublets, in fact) of H-like O$^{7+}$ and C$^{5+}$ at the laboratory wavelengths of 18.97~\AA\ and 33.74~\AA , respectively.
In Fig.~\ref{Lyalpha} we show the spectrum around these lines, and the best-fit model. 
The broad troughs of these lines from $\sim 0$ to $-2200$~\kms\ demonstrate nicely the velocity distribution of the outflow of \ngc , as well as the intricate overlap between intrinsic absorption and emission lines.
The --2050\,\kms\ component is marginally resolved for C$^{5+}$ thanks to the superior resolving power of the gratings at longer wavelengths  ($\sim 600$~\kms\ vs. $\sim 1000$ \kms\  for O$^{7+}$).

The model also includes the photo-ionised emission component that is discussed in more detail below (Sec.~\ref{emission}).
Interestingly, the emission lines observed in \ngc\ do not seem to be at rest with respect to the host galaxy, but they are blue-shifted by --450\,\kms .
Unlike the narrow absorption profiles, the emission lines are moderately broad and parametrized in the model with a turbulent velocity of $v_{\rm turb} = 600$~\kms\ (FWHM $\approx$ 1400~\kms ).

Due to instrumental broadening, there is blending of the --650\,\kms\ and --950\,\kms\  absorption troughs, as well as with the emission at --450\,\kms . 
This makes the fitted parameters of these components  in the model somewhat degenerate. 
\citet{Blustin07} did not identify a --950\,\kms\ component, and in the present spectrum too it could be part of the --650\,\kms\ component  (or some in-between velocity), but see attempts to tie the two below.
We tend to believe these velocity components since they are consistent with the UV ones, and since they are obtained from a rather constraining global fit to the entire X-ray spectrum, and not just from these individual lines. 

\subsection{Ionisation and Column Density}
The parameters of the six absorption components, which are associated here with three different outflow velocities, are listed in Table~\ref{params}.
The statistical significance of each component is given in the last column, in terms of the increase of the C-statistic ($\Delta C$) when that component is omitted from the model.
The slowest --650 \kms\ velocity component has a broad ionisation distribution of $-0.6 \lesssim \log \xi \lesssim 2.0$ represented here by three ionisation components, with the column density slowly increasing with ionisation, and with a possible gap in thermally unstable $\xi$ values. 
A similar trend was observed in the 2004 observation \citep{Blustin07}, as well as in other Seyfert outflows \citep{Holczer07, Behar09, Fukumura10, Laha14}.
An elaborate analysis of the Absorption Measure Distribution (AMD) 
is deferred to a following paper that will use a more sensitive ion-by-ion fit.

\begin{figure}
\begin{center}
\includegraphics[scale=0.35,angle=270]{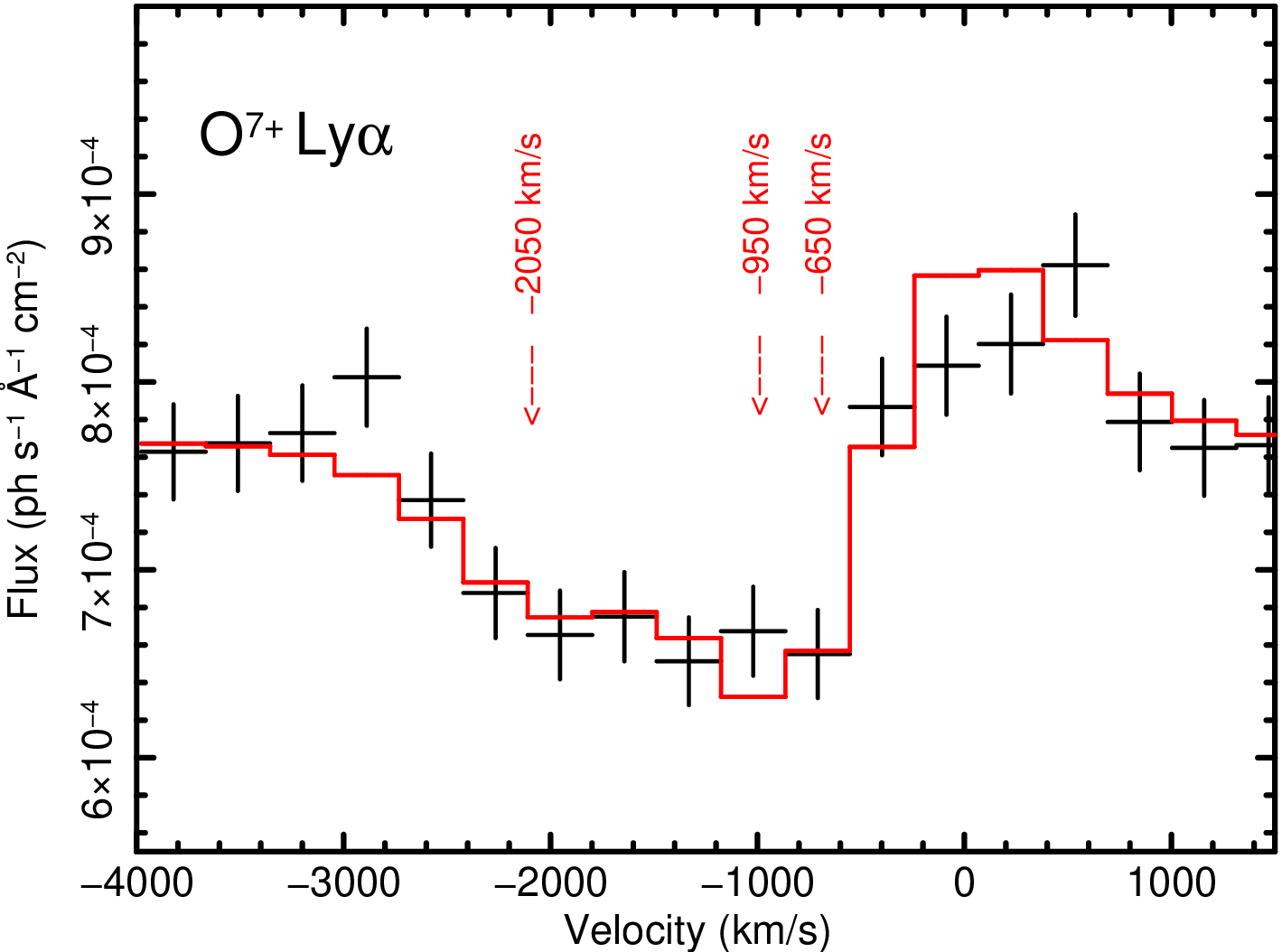} 
\includegraphics[scale=0.35,angle=270]{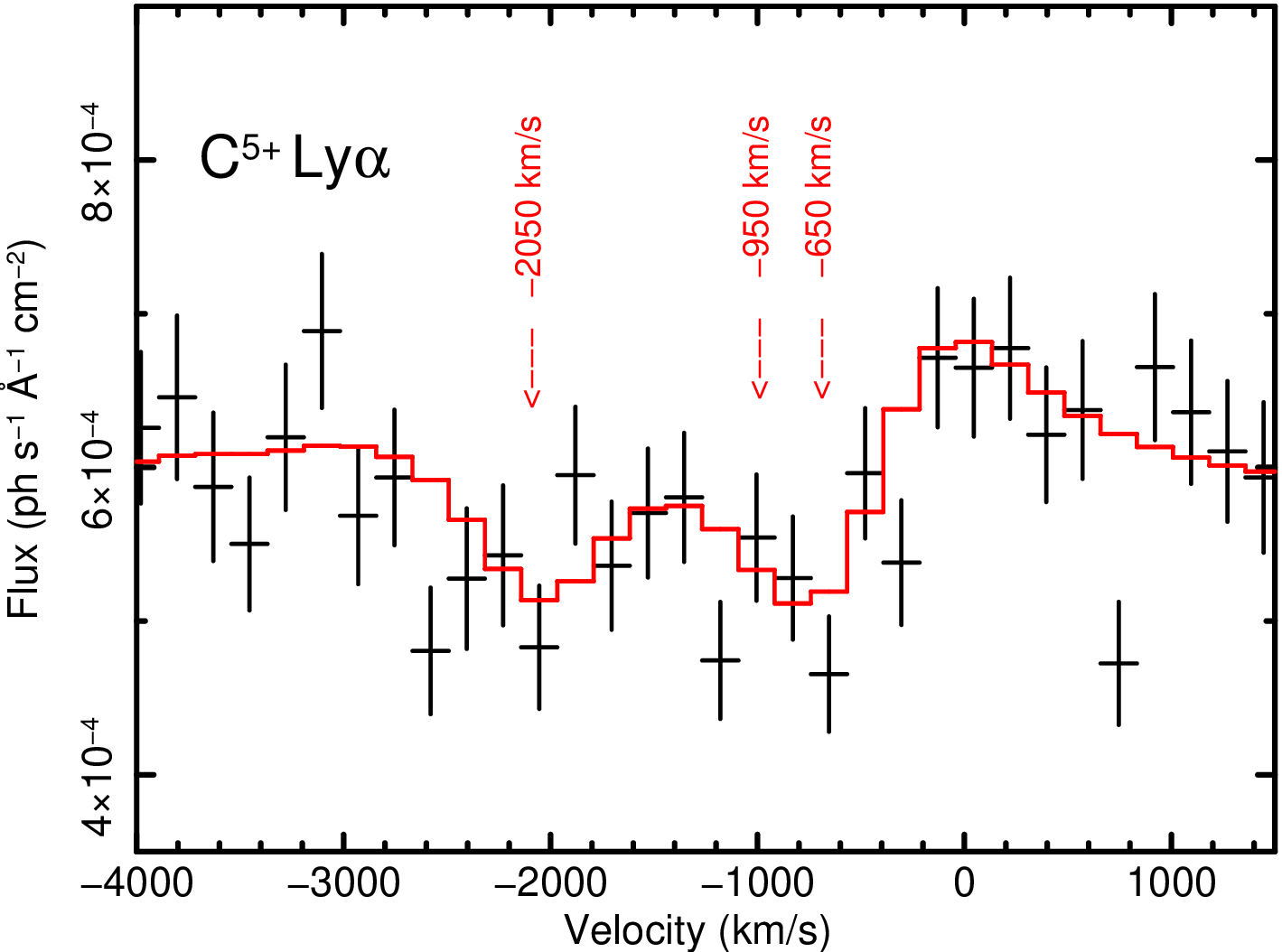} 
\end{center}
\caption{
Spectral region of the H-like O$^{7+}$ and C$^{5+}$ Ly$\alpha$ lines demonstrating the contribution of the three absorption components (marked on figure) and a broad (FWHM = 1400 \kms ) emission component (not marked) centered at --450\,\kms .
The instrumental spectral resolution is $\sim 1000$ \kms\ at the O$^{7+}$ line and $\sim 600$ \kms\ at the C$^{5+}$ line, as can be seen from the folded model (solid line). Note that the vertical flux scale does not go down to zero. 
}
\label{Lyalpha}
\end{figure}

The highest ionisation ($\log \xi = 2.7$) and highest column density ($2.2 \times 10^{21}$\,\cmsq) is found in the --950\,\kms\ component.
Its most notable feature is the Fe$^{19+}$ blend at 12.85 \AA , observed at 13.05~\AA\ (Fig.~\ref{zoom}),
which is one of the strongest absorption feature in the spectrum.
This component also accounts in the model for the higher charge states of Fe, and contributes to the central part of the Ly\,$\alpha$ O$^{7+}$ trough, although 
the \rgs\ can not truly resolve it from the --650\,\kms\ component (
see Fig.~\ref{Lyalpha}).
The counterpart --857~\kms\ component in the UV is observed only in Ly\,$\alpha$, which also indicates high ionisation, lending support to the connection between these two X-ray and UV components.
The high-ionisation at --950~\kms\ may also be, in fact, the extension of the broad ionisation distribution of the --650~\kms\ component, as its column density continues to increase with $\xi$.
When forcing all four ionisation components to share the same outflow velocity, the model for the entire spectrum converges to a velocity of --830\,\kms , since much of the column is in the --950\,\kms component, but the fit worsens by $\Delta C \approx 30$.

The fastest component (--2050\,\kms ) has one main ionisation component with $\log \xi = 2.0 \pm\ 0.3$, and a low column density of $\approx 10^{20}$~\cmsq .
A second, marginally significant low-ionisation component can be added at this velocity.
It can be seen, e.g., to capture the blue side of the trough of the O$^{6+}$ resonance line.
It improves the fit by $\Delta C \approx 50$. 
The fitted model in \citet{Blustin07} indicated relatively low ionisation ($\log \xi = 0.8^{+0.4}_{-0.3}$) for this fast component.
However, we identify it clearly in the H-like Ly\,$\alpha$ lines as seen in Fig.~\ref{Lyalpha}.
A more careful analysis of changes in the absorption troughs between 2004 and 2015, to be reported in a separate paper, is required before determining whether the ionisation has changed between the two epochs.
The total column density in all of the present components adds up to $\approx 3\times 10^{21}$~\cmsq , which is similar to the result of the AMD analysis of the 2004 spectrum in \citet{Blustin07}.

\begin{table*}
\caption{Outflow Absorption Components in \ngc }              
\label{params}      
\centering                                      
\begin{tabular}{c c c c c r}          
\hline\hline                        
Comp. & $v_{\rm out}~^a$ & $v_{\rm turb}$ & $\log \xi$ & $N_{\rm H}$ & $\Delta C$\\    
\# & (\kms ) &(\kms ) & (\ionpar ) & ($10^{20}$ \cmsq ) \\[+0.1cm]
\hline                                   
    1 & --650 $\pm$ 50 & 70$\pm$10  & --0.6 $\pm$ 0.2        & 0.2 $\pm$ 0.1 & 33\\[+0.1cm]
    2 & \nodata & 70$\pm$10 & 1.4 $\pm$ 0.1  & 1.0 $\pm$ 0.3 & 221 \\[+0.1cm]
    3 & \nodata  & 70$\pm$10 & 2.0 $\pm$ 0.1 & 5.5 $\pm$ 1.0 & 1027 \\[+0.1cm]      
    4 & --950$^{+50}_{-100}$ & 35$\pm$20 & 2.7 $\pm$ 0.2   & 22 $\pm$ 10 & 383 \\[+0.1cm]
    5 & --2050$^{+50}_{-160}$ & 60$\pm$30 & 2.0 $\pm$ 0.3      & 1.1 $\pm$ 0.3 & 82 \\[+0.1cm]
    6 & \nodata & 60$\pm$30 & 0.3 $\pm$ 0.2      & 0.1 $\pm$ 0.1 & 48\\[+0.1cm]
\hline                                             
\multicolumn{6}{l}
{\footnotesize $^a$ velocities and widths of Components 1-3 and those of 5-6 are tied}\\
\end{tabular}
\end{table*}

\subsection{Elemental abundances}
\label{abundances}

We fixed the abundance of O to its solar value and fitted for the abundances of C, N, Ne, Mg, S, and Fe, all of which have significant features in the \rgs\ spectrum. Other elements were left at their solar values. 
For the sake of simplifying the model, we also tied the abundances of the emission components to those of the absorber. 
The abundances  we obtained are 1.6$\pm$0.4, 1.6$\pm$0.6, 1.8$\pm$0.5, 0.3$\pm$0.2, 0.8$\pm$0.3, and 1.4$\pm$0.4, for C, N, Ne, Mg, S, and Fe, respectively.
All abundances are given with respect to the solar values of \citet{Asplund09}, and in effect represent $A_{\rm Z}/A_{\rm O}$ abundance ratios (and not absolute abundances $A_{\rm Z}/A_{\rm H}$).
If the abundance ratio $A_{\rm O}/A_{\rm H}$  is solar, the results seem to have a slight tendency for super-solar abundances, 
except for Mg and S. 
Note that the Mg lines around 9\,\AA\ and the S lines of S$^{13+, 12+, 11+}$ above 30\,\AA , are in regions where the \rgs\ is less sensitive.

\subsection{Emission features}
\label{emission}

There are several conspicuous emission lines in the spectrum, most notably the H-like Ly\,$\alpha$ lines of O$^{7+}$ and C$^{5+}$ (observed, respectively, at $\sim$ 19.3 and 34.3~\AA ), the so-called He-like triplet of O$^{6+}$ and Ne$^{8+}$, the strongest of which are the forbidden lines, 
indicative of recombination line driving, and photo-ionisation conditions. 
All three O$^{6+}$ lines are apparent in the spectrum around 22~\AA , while only the forbidden line of Ne$^{8+}$ around 13.9 \AA\ can be discerned due to blending with Fe absorption lines in that region. Bright narrow radiative recombination continua (RRCs) at 32.1\,\AA\ (C$^{4+}$), 25.7\,\AA\ (C$^{5+}$), 22.8\,\AA\ (N$^{5+}$), 17.0\,\AA\ (O$^{6+}$), and 14.4\,\AA\  (O$^{7+}$), can also be identified in the spectrum. All wavelengths are quoted in the observed, redshifted frame of Figs.~\ref{zoom} and \ref{zoom2}. 

We are able to model all of these emission features with two ionisation ({\it photemis}) components, whose details are listed in Table~\ref{em_comp} (and which also includes the hot plasma component of the next section). 
The lines require non-thermal broadening, which is parametrized in the model with a turbulent velocity of $v_{\rm turb} = 600$~\kms\ that probably represents a distribution of projections from the ionisation cone, rather than actual turbulence.
The ionisation parameters of the two emission components are $\log \xi = -1$ and 1 \ionpar , each contributing to the spectrum as can be seen by their statistical significance $\Delta C$ (when omitted) in the last column of Table~\ref{em_comp}.
This likely represents a range of $\xi$ values. 
It is reminiscent of the range of the absorption components, but excluding the highest Fe-L charge states, whose line emission is less efficient in recombination (See Sec. \ref{hotgas}).
The abundances of these emission components, therefore, were fixed to those of the absorber, assuming they originate from the same body of gas.

\begin{table}
\caption{Model Emission Components}              
\label{em_comp}      
\centering                                      
\begin{tabular}{c c c c r}          
\hline\hline                        
Model & $v_{\rm out}~^a$ & $\log \xi$ & $EM$ & $\Delta C$\\    
Component & (\kms ) & (\ionpar ) &  ($10^{63}$\,cm$^{-3}$) &  \\[+0.1cm]
\hline                                   
    {\it photemis} & --470 $\pm$ 150 & --1.0 $\pm$ 0.2        &  & 111\\[+0.1cm]
    {\it photemis} & \nodata & 1.0    $\pm$ 0.2      &  & 270 \\[+0.1cm]
\hline                                   
&& $kT$ (keV)& &\\[+0.1cm]
\hline                                   
    {\it APEC}  & --250 $\pm$ 120 & 0.35 $\pm$ 0.03 & 4.0$\pm$1.5 & 65\\[+0.1cm]      
\hline                                             
\multicolumn{5}{l}
{\footnotesize $^a$ {\it photemis} velocities tied} \\
\end{tabular}
\end{table}

\subsection{Emission from star formation}
\label{hotgas}

\begin{figure}
\includegraphics[scale=0.35,angle=270]{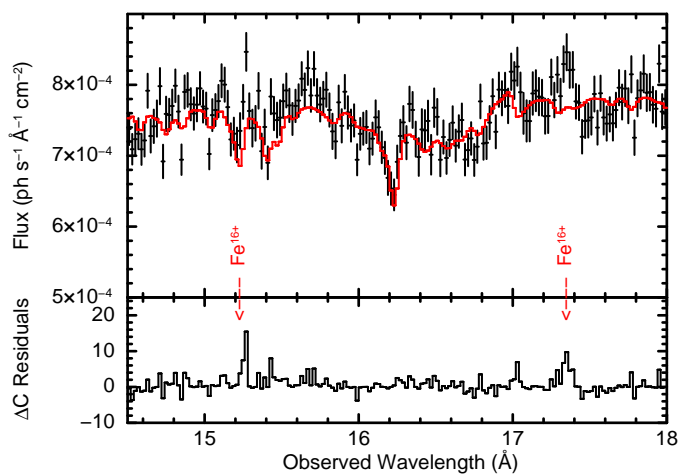} 
\caption{
The spectrum of \ngc\ with a model that does not include the collisionally-ionised plasma component.
Residuals of the Ne-like Fe$^{16+}$ 2p - 3d line at 15.01 and 2p - 3s line at 17.1~\AA\ (cosmologically redshifted in the figure) demonstrate the presence of collisionally-ionised plasma, likely originating in the circumnuclear star forming ring of \ngc .}
\label{coll}
\end{figure}

After fitting two emission components from photo-ionised gas, we are still left with weak yet significant residuals in the Fe$^{16+}$ line at 15.25\,\AA\ and the 17.38\,\AA\ doublet (both observed frame), which are shown in Fig.~\ref{coll}. 
The L-shell lines in photo-ionised plasmas, where they form by means of recombination, are weak relative to K-shell lines \citep{Kallman96, Kinkhabwala02}.
The lower cosmic abundance of Fe compared to C, N, and O also increases this effect, although in collisionally-ionised plasmas the Fe-L resonance lines have high collision strengths that compensate for the lower abundance, and are therefore prominent in astrophysical spectra \citep[e.g.,][]{Behar01}. 
We therefore interpret these lines as arising from the star formation ring in the center of \ngc\ \citep{Diaz-Santos07}.

Adding a collisionally-ionised component \citep[\small{APEC,}][]{Smith01} provides a good fit to these lines.
Fig.~\ref{coll} can be compared with that spectral region in Fig.~\ref{zoom}, where the model includes the collisional component.
The turbulent velocity broadening can not be constrained and is fixed at $v_{\rm turb} = 100$~\kms .
As in the photo-ionised components, this broadening may represent a kinematic spread of velocities projected along the line of sight, rather than actual turbulence.
The best-fit temperature corresponds to 0.35 $\pm$ 0.03~keV.
The Fe-L lines are slightly blueshifted by --250 $\pm$ 120 \kms . 
Although only marginally indicative of an outflow, this component can be associated with a starburst wind.
This component also produces some flux in O$^{7+}$ that blends with the emission by the photo-ionised gas.
Since only the Fe$^{16+}$ lines are unambiguously ascribed to the collisional gas, we are not able to fit for its abundances, and not even for its overall metallicity.
We therefore tie its metallicity to the Fe abundance of the photo-ionised components, which is fitted to be $1.4 \pm\ 0.4$ with respect to O, and in solar units (see Sec.~\ref{abundances}).

The X-ray flux implied by the collisionally-ionised plasma model in the \rgs\ band is $1.4\times 10^{-13}$\,\flux , which corresponds to a luminosity of $L_{\rm X} = 7\times 10^{40}$\,\ergs . 
Star-formation X-ray luminosity tends to correlate with the Far IR luminosity $L_{\rm FIR} \sim 10^{3.7}L_{\rm X}$ \citep{David92, Ranalli03}.
In \ngc\ $L_{\rm FIR} = 7\times 10^{44}$\,\ergs\ \citep[][corrected for distance]{David92}, so that $L_{\rm FIR} / L_{\rm X} = 10^4$, which is just between the two archetypical, nearby starburst galaxies {\small M\,82}, and {\small NGC\,253}, for which $L_{\rm FIR} / L_{\rm X} = 0.4\times 10^4$ and $1.8\times 10^4$, respectively \citep{David92}.
Thus, the X-rays emitted by the star-formation in \ngc\ are dominated by the AGN ($L_{\rm X} = 2\times 10^{43}$\,\ergs), but are exactly as expected from the FIR star-formation emission.

Using an angular distance of 63.2\,Mpc, the collisional component yields an emission measure ($EM = n_{\rm e}n_{\rm H}V$, density square times volume) of $4\times 10^{63}$\,cm$^{-3}$.
This is more than ten times lower than that of the photo-ionised gas (see Sec.~\ref{location} below), but much higher than $7.5\times 10^{61}$\,cm$^{-3}$, that was measured for {\small NGC\,253} \citep{Pietsch01}.
This result is consistent with the similarly higher FIR and X-ray luminosities in \ngc , and manifests its higher star formation rate than {\small M\,82} and {\small NGC\,253}. 
For a typical interstellar density of 1\,cm$^{-3}$, the above $EM$ corresponds to a volume of approximately 1\,kpc$^3$, which nicely matches the 1\,kpc diameter of the circum-nuclear star forming ring in \ngc\ \citep{Diaz-Santos07}. 

\subsection{Galactic absorption}
For the neutral absorption through our galaxy, we use a column density of $5.5 \times 10^{20}$~\cmsq\ that includes both HI and H$_2$ (see end of Sec.~\ref{intro}).
We use the {\it tbabs} model \citep{Wilms00} in {\small Xspec} with solar abundances, as its position of the neutral O K-shell edge at about 23.1\,\AA\ better describes the spectrum of \ngc\ than the 23.5\,\AA\ edge in other models (see Fig.~\ref{zoom2}).
The newer model {\it tbnew} brings the edge to slightly shorter wavelengths, due to absorption lines, but does not improve the fit.  
Since {\it tbabs} does not include absorption lines, we added the K$\alpha$ absorption lines of neutral O and N to the model at 23.5\,\AA\ and 31.3\,\AA, respectively. These are prominent in the spectrum (Fig.~\ref{zoom2}), and do not suffer from blending with the AGN absorber \citep[as opposed to {\small Mrk\,509},][]{Kaastra11, Pinto12}.
The exact same method was used successfully by \citet{Reeves16} for the X-ray spectrum of {\small Ark\,120}.

It is puzzling that the fit prefers ($\Delta C = 350$) a much lower column of $3.4 \times 10^{20}$~\cmsq\ than the nominal $5.5 \times 10^{20}$~\cmsq .
The improvement is especially seen at longer wavelengths (>20\,\AA ), and is likely driven by the N and C edges.
This could indicate non-solar ISM abundances, calibration issues, or an inappropriate continuum model. 
Attempts to add a soft excess and then to absorb it did not improve the global fit. 
We fitted the archival 2004 \rgs\ spectrum of \ngc\ to find that, unlike here, the higher neutral column is obtained as expected.
We defer a more detailed study of the local, neutral absorption to a combined X-ray and UV analysis, as was done for {\small Mrk\,509} \citep{Pinto12}.

We identify in the spectrum a strong He-like O$^{6+}$ absorption line at 21.6\,\AA , which is ascribed to hot gas in the Galactic halo, or in the local group. 
The H-like O$^{7+}$ Ly\,$\alpha$ line at 18.97\,\AA\ is marginally present, just next to the (redshifted) O$^{6+}$ He\,$\beta$ line of the \ngc\ outflow. 
We thus add a collisionally-ionised {\small XSTAR} absorption component ({\it hotabs}) to represent this local ($z = 0$) absorber.
The improvement to the global fit is $\Delta C = 65$ for over 1500 degrees of freedom. 
The fitted parameters of this component are a temperature of $1.5\times 10^6$\,K and a H column density of $2 \times 10^{19}$~\cmsq , under the assumption of solar abundances, which adequately accounts for the weak observed features.

\subsection{Unidentified lines}
\label{unidentified}

The \rgs\ spectrum has a bonafide absorption line at $28.56 \pm\ 0.02$\,\AA\ (see Fig.~\ref{zoom2}), which corresponds to $\sim$28.10\,\AA\ in the rest frame of \ngc . 
There is no obvious identification at these wavelengths. 
The nearby C$^{5+}$ Ly\,$\beta$ line at 28.466\,\AA , would need to be approaching at --3800\,\kms\ in \ngc , or receding by 1000\,\kms\ in our frame to produce the observed feature, neither velocity of which is observed in other lines, including not in C$^{5+}$ Ly\,$\alpha$.
A closer line is the 2p - 4d line of Ar$^{10+}$ at  28.529\,\AA , but this ion would also have a stronger 2p -3d line at 34.328\,\AA\ \citep{Lepson03}, which is not observed.
There is no prominent line that we know of close to 28.10\,\AA .

The \rgs\ spectrum also features two bright emission lines observed around 32.6\,\AA\ and 33.4\,\AA , which can be seen rising high above the best fit model in Fig.~\ref{zoom2}.
To measure their positions and fluxes, we fitted the lines with two gaussians. 
The line widths indicate they are unresolved.
The best fitted positions are $32.60 \pm\ 0.02$ \AA\ and $33.37 \pm\ 0.02$~\AA , or $32.08 \pm\ 0.02$ \AA\ and $32.84 \pm\ 0.02$ in the rest frame of \ngc .
The fluxes (corrected for Galactic absorption) are $(4.3 \pm\ 1.0) \times 10^{-5}$ and $(2.4 \pm\ 0.7) \times 10^{-5}$~\flux , respectively. 

We struggle to provide convincing interpretations for these emission lines.
The only bright lines in this region of the spectrum are the K-shell lines of C, and in particular the He\,$\gamma$ and He\,$\delta$ lines of C$^{4+}$ at 32.754 \AA\ and 33.426 \AA .
If ascribed to  C$^{4+}$, the observed lines are blue-shifted by $\sim 160$ and 50~m\AA ,
but narrow emission lines in the gratings can only arise from a point source along the line of sight towards \ngc , so this is ruled out.
We could not identify possible emission lines that would occur at 32.08\,\AA\ and at 32.84\,\AA\ in the rest frame of \ngc .


\subsection{Location and physical state of the outflow}
\label{location}
One of the main goals of the present campaign is to track down the position of the outflow in order to determine its mass outflow rate and kinetic power, and thereby its possible influence on the host galaxy.
The major attempt will be made in a separate paper discussing the variability of the absorber, which is subtle.
Here, we limit ourselves to simple considerations.
Most generally, the column density $N_{\rm H}$ as deduced from a single ion that forms at an ionisation parameter $\xi$, can be written as

\begin{equation}
N_{\rm H}=\int_{r_{min}}^{r_{max}}{n_{\rm H}dr}=\frac{L}{\xi}\left(
\frac{1}{r_{min}}- \frac{1}{r_{max}}\right)
\end{equation}

\noindent
If $r_{\rm max} >> r_{\rm min}$ then 

\begin{equation}
r_{\rm min} \approx \left(\frac{1}{r_{\rm min}} - \frac{1}{r_{\rm max}}  \right)^{-1} = \frac{L}{\xi N_{\rm H}}
\label{rmin}
\end{equation}

\noindent 
Given an ionising X-ray luminosity of $2\times 10^{43}$\,\ergs\ and a total column density of $3\times 10^{21}$\,\cmsq ,  
Eq.~\ref{rmin} would put the present highest ionisation component of $\xi = 10^3$ ($r_{\rm min} \approx$) a few pc away from the central source, while the lower ionisation components, where most of the mass is, are as far as 10 {\it kpc} away.
This is in line with the present 2015 absorption spectrum being essentially unchanged from that of 2004 \citep{Blustin07}. 

Since we observe a wide range of $\xi$ values, Eq.~\ref{rmin} clearly can not tell the whole story of the outflow.
Whewell et al. (2015) recently used this estimate for {\small NGC\,5548} and found $r_{\rm min} = 13.9 \pm\ 0.6$~pc.
On the other hand, an estimate based on changes in UV absorption troughs in {\small NGC\,5548},
assuming they are only due to varying ionising flux over 16 years, yields a distance of 3.5\,$\pm$\,1.0\,pc for one component, 5 - 70 pc for three other components and two more beyond 100~pc  \citep{Arav15}. 
This suggests that $r_{\rm max} >> r_{\rm min}$ may not apply, and thus, the above estimate for \ngc\ may also be too high.

Another way to obtain an idea of the location and density of the absorber is to associate the narrow line emitting gas with the outflow. This notion is corroborated here by the actual blueshift of the emission lines, and by the overlap of $\xi$ values. 
An emission-line photon flux $F_{ji}$  can be related to the volume integrated emission measure:

\begin{equation}
EM=\int_V{n_{\rm e}n_{\rm H}dV}=\frac{4\pi d^2F_{ji}}{f_{q+1}A_ZP_{ji}}
\end{equation}

\noindent where $d$ is the distance to the source.
The fraction of ions in the charge state of the line $q$ is denoted by $f_q$ (a function of $\xi$ typically peaks at $\sim$0.5). 
$A_Z$ is the elemental abundance with respect to H, and $P_{ji}$ is the line power (in units of cm$^3$s$^{-1}$).
The $EM$ of the brightest line in the spectrum of \ngc , namely that of  the forbidden line of O$^{6+}$, is $\sim 10^{65}$\,cm$^{-3}$. 
In the approximation of $r_{\rm max} >> r_{\rm min}$, the $EM$ can be used to estimate the opening angle of the ionisation cone \citep[see Eq.~4 in][]{Behar03}, but here this estimate yields an unphysical value that is above $4\pi$, giving yet another indication that this approximation is not valid here.

On the other hand, if the absorber size $\Delta r$ is much smaller than its distance from the source $r$, i.e., a cloudlet, and assuming $n_{\rm H} \approx n_{\rm e} = n$ one can approximate $EM \approx n^2\Delta r^3$ and $N_{\rm H} \approx n\Delta r$, which can be directly solved to yield 

\begin{equation}
n=N_{\rm H}^3/EM 
\label{density}
\end{equation}

\noindent Using the above values for O$^{6+}$, Eq.~\ref{density} yields $n = 0.3$\,cm$^{-3}$.
Since O$^{6+}$ forms at about $\xi = 30$ this implies a distance from the center of $r = 300$\,pc;
a result to be taken with caution, as a uniform cloudlet is clearly over simplistic.
For one, a strong density gradient would need to be present inside to explain the large range of $\xi$ observed
\citep[see, e.g.,][]{Rozanska06, Stern14}.

\section{Discussion and Conclusions}
\label{discussion}

We showed that the high-resolution X-ray spectrum of \ngc\ requires a multitude of components that include beyond neutral and ionised absorption at $z = 0$, also several components of ionised absorption in the outflow of the AGN, low and high photo-ionisation emission components, as well as X-ray lines from the circumnuclear starburst of \ngc , and its wind.
The main results of this analysis can be summarized as follows:

\begin{itemize}
        \item{The outflow velocities of --650, --950, and --2050~\kms\ are consistent with the three components detected simultaneously in the UV, although the presence of the --950~\kms\ component is not unambiguous in the X-ray spectrum, and it  could also be associated with an extension to high ionisation of the --650~\kms\ component.}\\
        \item{The absorber of \ngc\ shows a broad ionisation distribution at least in the slowest velocity component. This is evidently a ubiquitous feature of Seyfert outflows. }\\
          \item{Preliminary estimates of the position of the absorber indicate it is a few pc away from the center or more, but a careful variability study in the X-rays and the UV will provide better constraints.}\\
        \item{The narrow emission lines from photo-ionised gas are blue-shifted, indicating that they may originate from the absorbing outflow. This X-ray narrow line region is likely the same extended region observed in the UV and resolved in nearby Seyfert galaxies \citep[e.g.,][]{Ruiz05}. }\\
          \item{Subtle but significant emission lines are detected from the hot circumnuclear starburst ring of \ngc . The high flux and emission measure of this plasma indicates star formation that exceeds nearby starbursts. }\\

\end{itemize}

Future work will seek and analyse small changes in absorption over the course of the months-long 2015 campaign, and in comparison with the 2004 observation. X-ray absorption variability will be compared with a similar analysis of the UV HST/COS spectrum of \ngc , where the S/N ratio is much better. 
The goal is to identify a change in the X-ray and UV absorbers that can be ascribed to changes in the ionising continuum, and thus to obtain an idea of the density, distance from the center, and in turn the mass outflow rate.

\begin{acknowledgements} 
This work is supported by NASA grant NNX16AC07G through the \xmm\ Guest Observing Program, and through grants for HST program number 14054 from the Space Telescope Science Institute, which is operated by the Association of Universities for Research in Astronomy, Incorporated, under NASA contract NAS5-26555. 
The research at the Technion is supported by the I-CORE program of the Planning and Budgeting Committee (grant number 1937/12). 
NA is grateful for a visiting-professor fellowship at the Technion, granted by the Lady Davis Trust.
EB received funding from the European Union's Horizon 2020 research and innovation programme under the Marie Sklodowska-Curie grant agreement no. 655324. 
SB and MC acknowledge financial support from the Italian Space Agency under grant ASI-INAF I/037/12/0.
POP and FU acknowledge support from CNES and from PNHE of CNRS/INSU.
GP acknowledges support by the Bundesministerium f\"{u}r Wirtschaft und 
Technologie/Deutsches Zentrum f\"{u}r Luft- und Raumfahrt 
(BMWI/DLR, FKZ 50 OR 1408 and FKZ 50 OR 1604) and the Max Planck Society. We thank the referee for useful comments that have been implemented in the manuscript.
 \end{acknowledgements}

\bibliographystyle{aa}

\end{document}